\RequirePackage{ifpdf}
\documentclass[12pt,letterpaper]{article}
\pdfoutput=1
\usepackage{jheppub}
\usepackage{epsfig}
\usepackage{enumitem}
\usepackage[latin1]{inputenc}
\usepackage{bbm,amsfonts}
\usepackage{rotating,graphicx}
\usepackage{amssymb,amsmath,amsfonts,mathtools}
\usepackage{fancybox,diagbox}
\usepackage{enumerate}
\usepackage{dsfont}
\usepackage{verbatim}
\usepackage{wrapfig}
\usepackage{slashed}
\usepackage{shuffle}
\usepackage{subcaption}
\usepackage{braket}
\usepackage{pdflscape}
\usepackage{accents}
\usepackage{afterpage}

\author{Marco S. Bianchi}
 
\affiliation{Instituto de Ciencias F\'isicas y Matem\'aticas, Universidad Austral de Chile, Casilla 567, Valdivia, Chile}

\emailAdd{marco.bianchi@uach.cl}  
\title{Protected and uniformly transcendental} 

\abstract{
We show that the two-point function of protected bi-scalar operators in ${\cal N}=4$ SYM evaluated in dimensional regularization exhibits a uniform degree of transcendentality up to three-loop order. We conjecture that this property holds for the whole perturbative series and leverage the explicit results to postulate a prediction for the leading, order $\epsilon$, correction to all loop orders. We also consider the soft limit of three-point functions of such operators in momentum space and point out a simple and surprising perturbative relation to two-point functions, which we also extrapolate to all loop orders.
}

\def\Tr{\textrm{Tr}}

\newtheorem{conjecture}{Conjecture}

\numberwithin{equation}{section}

\newlength{\dhatheight}

\begin{document}

\maketitle
\allowdisplaybreaks

\section{Introduction}

In this note we focus on the perturbative calculation of two-point functions of protected scalar bi-linear operators in $\mathcal{N}=4$ SYM of the form $O_0=\Tr\left(XX\right)$, where $X$ is one of the theory's complex scalars. Such correlators are tree-level exact. So are their three-point functions \cite{Intriligator:1999ff}.
Consistently, computing the former in dimensional regularization with $d=4-2\epsilon$, a subleading in the regulator $\epsilon$ result appears for the operators' norm at quantum level.

Such a computation is seldom required in ${\cal N}=4$ SYM perturbation theory.
The context in which it originated was that of three-point functions, and the extraction of their structure constants via a calculation in dimensional regularization \cite{Bianchi:2023cbc}. 
In order to perform such a procedure one needs to consider a basis of renormalized operators, whose two-point functions do not mix. It is also reasonable to normalize the final result by the norms of the resulting conformal primary operators, to extract universal structure constants. Both tasks require a knowledge of the two-point functions.

Since three-point functions are generically divergent before renormalization, $\epsilon$ subleading corrections of two-point functions at a given loop can alter the finite contributions at higher loop level, hence the necessity of determining their $\epsilon$ expansion up to the required order. In fact, taking into account such contributions was crucial in order to recover previously computed results in \cite{Bianchi:2018zal,Bianchi:2019jpy,Bianchi:2022oyz,Bianchi:2023cbc}.

Determining the $\epsilon$ expansion of two-point functions including $\epsilon$ subleading terms in dimensional regularization is actually more relevant for un-protected operators, but while we were computing them we also considered the case of protected ones, as a consistency check.
From \cite{Bianchi:2023cbc}, to the needed order in $\epsilon$ for three-loop structure constants plus one additional unit for safety, the result for the quantum corrections to the two-point function of twist-two scalars in momentum space reads
\begin{align}
&\left\langle O_{0}(p) O_{0}(-p) \right\rangle = \frac{N^2-1}{8\pi^2} \bigg(
\left(\tfrac{1}{\epsilon }+2+\left(4-\tfrac{1}{2}\zeta _2\right) \epsilon +\left(8-\zeta _2-\tfrac{7}{3} \zeta _3\right) \epsilon ^2 \right.\nonumber\\&\left.~~~~
+\left(16-2 \zeta _2-\tfrac{14}{3} \zeta _3-\tfrac{47}{16} \zeta _4\right) \epsilon ^3
+ {\cal O}\left(\epsilon ^4\right)\right)
\nonumber\\&~~~~
+\left(-12 \zeta _3+\left(-24 \zeta _3-18 \zeta _4\right) \epsilon 
+\left(12 \zeta _2 \zeta _3-48 \zeta _3-36 \zeta _4-84 \zeta _5\right) \epsilon ^2
+{\cal O}\left(\epsilon ^3\right)
\right)\lambda
\nonumber\\&~~~~
+
\left(
100 \zeta _5 +\left(244 \zeta _3^2+200 \zeta _5+250 \zeta _6\right) \epsilon +{\cal O}\left(\epsilon ^2\right)
\right)\lambda^2
\nonumber\\&~~~~
+
\left(
-980 \zeta_7+{\cal O}\left(\epsilon ^1\right)
\right)\lambda^3 + {\cal O}\left(\lambda^4\right)
\bigg)
\end{align}
with some conventions and regularization scheme, that we spell out in more detail below.
Normalizing by the tree level contribution gives
\begin{align}
\frac{\left\langle O_{0}(p) O_{0}(-p) \right\rangle}{\left\langle O_{0}(p) O_{0}(-p) \right\rangle^{(0)}} &= 1+ \left(-12 \zeta _3 \epsilon -18 \zeta _4 \epsilon ^2 + 6 \left(\zeta _2 \zeta _3-14 \zeta _5\right) \epsilon ^3+{\cal O}\left(\epsilon ^4\right) \right) \lambda
\nonumber\\&
+\left(100 \zeta _5 \epsilon + \left(244 \zeta _3^2+250 \zeta _6\right) \epsilon ^2+{\cal O}\left(\epsilon ^3\right) \right) \lambda ^2
\nonumber\\&
+ \left(-980 \zeta _7 \epsilon+{\cal O}\left(\epsilon ^2\right) \right) \lambda ^3 + {\cal O}\left(\lambda^4\right)
\end{align} 
Such a result hints at the fact that the quantum corrections to the two-point function may be expressible in terms of a factor containing multiple zeta values of uniform degree of transcendentality $2l$, upon the standard assignment of transcendentality to $\epsilon$. 
This was an unforeseen occurrence to us, although perhaps not so surprising as uniform transcendentality appears rather ubiquitously in ${\cal N}=4$ SYM.
The claim of uniform transcendentality is a bit of a stretch at the moment, but it is what motivated the following deeper analysis.

As mentioned, uniform transcendentality properties manifest in various sectors of $\mathcal{N}=4$ SYM.
In the very context of twist-two operators, spinning unprotected ones exhibit anomalous dimensions expressed in terms of harmonic sums with uniform degree from seminal results \cite{Kotikov:2003fb,Kotikov:2004er} to higher loop findings \cite{Marboe:2014sya,Marboe:2016igj,Kniehl:2021ysp}. Their OPE coefficients with two protected operators also satisfy such a property \cite{Eden:2012rr}. 
For partially on-shell objects such as form factors, uniform transcendentality also manifests both in numeric quantities such as for the Sudakov form factor \cite{vanNeerven:1985ja,Gehrmann:2011xn,Huber:2019fxe,Agarwal:2021zft,Lee:2021lkc} and at the functional level in the remainder function of multi-point form factors \cite{Bork:2010wf,Brandhuber:2012vm,Brandhuber:2014ica,Banerjee:2016kri,Lin:2020dyj}.
Finally, uniform transcendentality is also ubiquitous in scattering amplitudes in ${\cal N}=4$ SYM theory \cite{Bern:2005iz,DelDuca:2009au,DelDuca:2010zg,Goncharov:2010jf,Arkani-Hamed:2012zlh}, where they can be expressed in terms of 
polylogarithms of uniform transcendental degree.

In this note we present another instance, two-point functions of protected bi-scalars, where uniform transcendentality kicks in. To the best of our knowledge such a result is original.

Besides being interesting and aesthetically satisfactory, uniform transcendentality is a powerful tool. 
It streamlines the evaluation of Feynman integrals, properly organizing the differential equations they satisfy \cite{Henn:2013pwa}. It also plays a crucial role in bootstrap methods for the computation of amplitudes \cite{Dixon:2011pw,Dixon:2011nj,Dixon:2013eka,Drummond:2014ffa,Caron-Huot:2016owq,Dixon:2016nkn,Drummond:2018caf,Caron-Huot:2019vjl,Dixon:2020cnr} and form factors \cite{Brandhuber:2012vm,Dixon:2020bbt,Dixon:2022rse,Dixon:2022xqh}, where a basis of suitable functions in conjured and their coefficients are fixed via all available physical requirements.

More similar to the problem we face in this note, extracting suitable factors from propagator integrals in such a way to expose a uniformly transcendental expansion helps reconstructing analytical results from numerics via PSLQ, as in \cite{Lee:2011jt}, whose explicit results we abundantly rely on in this calculation. Restricting the basis of possible transcendental numbers entering the expansion, to those obeying uniform transcendentality, offers a more efficient reconstruction.

Aside from probing the uniform transcendentality of the given two point-functions and conjecturing its extension to all loops, we propose a general expression fot the leading term in the $\epsilon$ expansion to all loops.
Furthermore, we consider the soft limit of three-point functions of the same operators in momentum space, an object relevant for the extraction of structure constants \cite{Bianchi:2023cbc}, and discover a simple relation connecting it to two-point functions, which we also conjecture to hold to all loops.

\section{Two-point functions at higher order in the regulator}

In order to investigate if a transcendentality pattern emerges, we computed the two-point function of scalar twist-two operators in ${\cal N}=4$ SYM with gauge group $SU(N)$ up to three-loop order in perturbation theory in the 't Hooft coupling $\lambda=\frac{g^2N}{16\pi^2}$ and to higher order in the $\epsilon$ expansion. We performed the computation exactly, in terms of master integrals, then expanded results around $d=4$.

The calculation is carried out along the lines of \cite{Bianchi:2023cbc}.
Summarizing, the relevant Feynman graphs are generated by QGRAF \cite{Nogueira:1991ex}, then processed with some routines in the computer algebra Form \cite{Vermaseren:2000nd,Ruijl:2017dtg} and reduced to master integrals via integration-by-parts (IBP) identities \cite{Chetyrkin:1981qh,Tkachov:1981wb} with Forcer \cite{Ruijl:2017cxj}, within the same framework.
The master integrals we use are the basis chosen in Forcer. They can be translated to other master integrals bases, such as the four-loop integrals of \cite{Baikov:2010hf,Lee:2011jt}, by inverting the relevant linear system of IBP reductions.
Dimensional regularization is enforced throughout and dimensional reduction scheme \cite{Siegel:1979wq} is applied in treating the diagrams algebra.
Finally, the $\epsilon$ expansions of the relevant master integrals, up to the maximal order of \cite{Lee:2011jt}, are substituted in the result.
The computation is performed and presented in momentum space where, at tree, one-, two- and three-loop level, integrals have one, two, three and four loops, respectively.

In order to expose uniform transcendentality, we factor out the tree level result
\begin{equation}\label{eq:perturbative}
\left\langle O_{0}(p) O_{0}(-p) \right\rangle = 2(N^2-1)\, \raisebox{-2.5mm}{\includegraphics[scale=0.1]{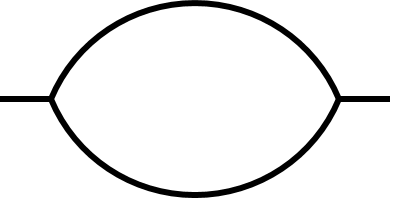}}\,\left( 1+n^{(1)}\lambda+n^{(2)}\lambda^2+n^{(3)}\lambda^3 + {\cal O}(\lambda^4) \right)
\end{equation}
where the picture represents the one-loop momentum integral
\begin{equation}
\raisebox{-2.5mm}{\includegraphics[scale=0.1]{M0L.png}} = \int \frac{d^{4-2\epsilon}k}{(2\pi)^{4-2\epsilon}}\frac{1}{k^2(k-p)^2} = \frac{\Gamma (1-\epsilon )^2 \Gamma (\epsilon )}{(4\pi)^{2-\epsilon}\Gamma (2-2 \epsilon )}
\end{equation}
and $n^{(l)}$ stands for the loop corrections normalized by the tree level bubble.
For simplicity we set $p^2=1$, the momentum dependence being trivially dictated by dimensional analysis.
For cleaner results, we also absorb extra factors $e^{\gamma_E \epsilon }$ and $(4\pi)^\epsilon$ in the definition of $\lambda$. 
In terms of master integrals, the one-loop correction reads
\begin{equation}
\left\langle O_{0}(p) O_{0}(-p) \right\rangle^{(1)} = \left(4-\tfrac{2}{\epsilon }\right) \raisebox{-0.8mm}{\includegraphics[scale=0.1]{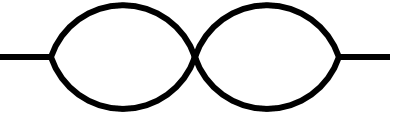}}+4\left(\tfrac{1}{\epsilon }-3\right)
\raisebox{-2.5mm}{\includegraphics[scale=0.1]{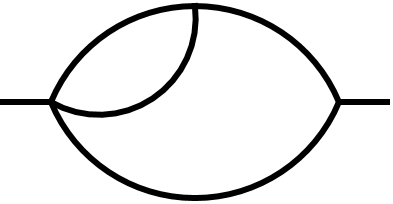}}
\end{equation}
Normalizing and plugging in the exact expression for the integrals gives
\begin{equation}\label{eq:1loopexact}
n^{(1)} = \frac{\Gamma (1-\epsilon )}{\epsilon ^2\Gamma (\epsilon )} \left(\frac{\Gamma (1-\epsilon ) \Gamma (\epsilon )^2}{\Gamma (-2 \epsilon )}-\frac{4\Gamma (2-2 \epsilon ) \Gamma (-1+2 \epsilon)}{\Gamma (1-3 \epsilon )}\right)
\end{equation}
The first few orders are
\begin{align}\label{eq:1L}
&n^{(1)} = -12 \zeta _3 \epsilon-18 \zeta _4 \epsilon ^2+\left(6 \zeta _2 \zeta _3-84 \zeta _5\right) \epsilon ^3+\left(64 \zeta _3^2-\tfrac{657 }{4}\zeta _6\right) \epsilon ^4+\left(\tfrac{741}{4} \zeta _3 \zeta _4+42 \zeta _2 \zeta _5
\right.\nonumber\\&\left.
-588 \zeta _7\right) \epsilon ^5
+\left(-32 \zeta _2 \zeta _3^2+\tfrac{3872}{5} \zeta _5 \zeta _3-\tfrac{18285}{16} \zeta _8\right) \epsilon ^6
+\left(-\tfrac{566}{3} \zeta _3^3+\tfrac{46647}{32} \zeta _6 \zeta _3+\tfrac{22287 }{20}\zeta _4 \zeta _5
\right.\nonumber\\&\left.
+294 \zeta _2 \zeta _7-4260 \zeta _9\right) \epsilon ^7
+\left(-813 \zeta _4 \zeta _3^2-\tfrac{1936}{5} \zeta _2 \zeta _5 \zeta _3+\tfrac{35824 }{7}\zeta _7 \zeta _3+\tfrac{11424}{5} \zeta _5^2-\tfrac{4843593}{640} \zeta _{10}\right) \epsilon ^8
\nonumber\\&
+\left(\tfrac{283}{3} \zeta _2 \zeta _3^3-\tfrac{49702}{15} \zeta _5 \zeta _3^2+\tfrac{2493779}{256} \zeta _8 \zeta _3+\tfrac{1366341}{160} \zeta _5 \zeta _6+\tfrac{205683}{28} \zeta _4 \zeta _7+2130 \zeta _2 \zeta _9
\right.\nonumber\\&\left.
-31836 \zeta _{11}\right) \epsilon ^9+\left(\tfrac{10784}{27} \zeta _3^4-\tfrac{49421}{8} \zeta _6 \zeta _3^2-\tfrac{47524}{5} \zeta _4 \zeta _5 \zeta _3-\tfrac{17912}{7} \zeta _2 \zeta _7 \zeta _3+\tfrac{108544 }{3}\zeta _9 \zeta _3-\tfrac{5712}{5} \zeta _2 \zeta _5^2
\right.\nonumber\\&\left.
+\tfrac{149328}{5} \zeta _5 \zeta _7-\tfrac{71432982333}{1415168} \zeta _{12}\right) \epsilon ^{10} +{\cal O}\left(\epsilon ^{11}\right)
\end{align}
and display uniform transcendentality 2, as it is evident from expression \eqref{eq:1loopexact}, after counting pole orders in each term.
At two loops the result of the calculation is the following
\begin{align}
&\left\langle O_{0}(p) O_{0}(-p) \right\rangle^{(2)} = \tfrac{4(1-2 \epsilon )^2}{\epsilon ^2} \, \raisebox{-0.8mm}{\includegraphics[scale=0.1]{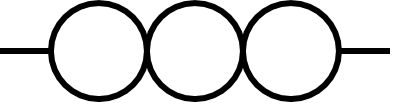}}
-\tfrac{24\left(6 \epsilon ^2-5 \epsilon +1\right)}{\epsilon ^2}\,\raisebox{-1.4mm}{\includegraphics[scale=0.1]{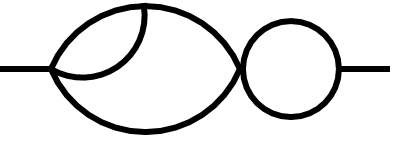}}
+\tfrac{16 \left(8 \epsilon ^2-6 \epsilon +1\right)}{\epsilon ^2}\,\raisebox{-2.5mm}{\includegraphics[scale=0.1]{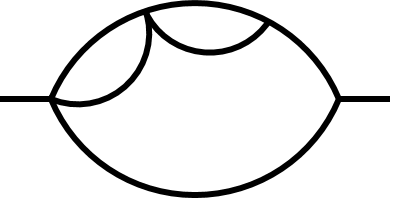}}\nonumber\\&
+\tfrac{16 (3 \epsilon -1) (4 \epsilon -1)}{\epsilon ^2}\,\raisebox{-2.5mm}{\includegraphics[scale=0.1]{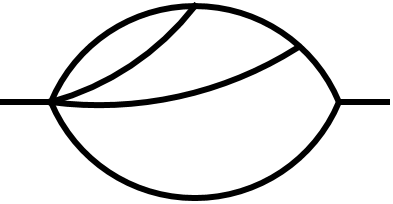}}
-\tfrac{12(2 \epsilon -1)}{\epsilon }\,\raisebox{-2.5mm}{\includegraphics[scale=0.1]{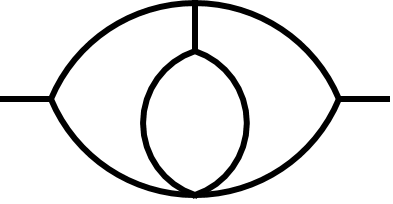}}\,
+\,\raisebox{-2.5mm}{\includegraphics[scale=0.1]{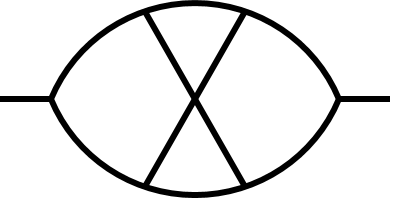}}
\end{align}
in terms of the master integrals depicted in the equation. 
Plugging in the exact expression for the three-loop integrals gives the two-loop correction
\begin{align}\label{eq:2L}
&n^{(2)} =100 \zeta _5 \epsilon+\left(244 \zeta _3^2+250 \zeta _6\right) \epsilon ^2+\left(732 \zeta _3 \zeta _4-100 \zeta _2 \zeta _5+1718 \zeta _7\right) \epsilon ^3+
    \left(\tfrac{7288}{3} \zeta _5 \zeta _3
\right.\nonumber\\&\left.    
    -\tfrac{1296}{5} \zeta_{5,3}-244 \zeta _2 \zeta _3^2+\tfrac{179647}{30} \zeta _8\right)\epsilon ^4 
    +\left(-\tfrac{10712}{3} \zeta _3^3+\tfrac{10717}{3} \zeta _6 \zeta _3+5763 \zeta _4 \zeta _5-1718 \zeta _2 \zeta _7
    \right.\nonumber\\&\left.
    +\tfrac{221140}{9} \zeta _9\right) \epsilon ^5
    +\left(\tfrac{1296}{5} \zeta _2 \zeta_{5,3}-\tfrac{16014}{7} \zeta_{7,3}-15641 \zeta _4 \zeta _3^2
    -\tfrac{7288}{3} \zeta _2 \zeta _5 \zeta _3
    +\tfrac{23420}{3} \zeta _7 \zeta _3
\right.\nonumber\\&\left.    
    +\tfrac{140021431}{1400} \zeta _{10}-\tfrac{4278 }{7}\zeta _5^2\right)\epsilon ^6 +
    \left(9504 \zeta _3 \zeta_{5,3}-\tfrac{58592}{5} \zeta_{5,3,3}+\tfrac{10712}{3} \zeta _2 \zeta _3^3-\tfrac{3124952}{45} \zeta _5 \zeta _3^2
    \right.\nonumber\\&\left.
    -\tfrac{836833}{18} \zeta _8 \zeta _3+\tfrac{164813}{4} \zeta _5 \zeta _6+\tfrac{276033}{10} \zeta _4 \zeta _7+\tfrac{36679873}{30} \zeta _{11}-\tfrac{4967092 }{9}\zeta _2 \zeta _9\right)\epsilon ^7 + {\cal O}(\epsilon^8)
\end{align}
which manifestly displays uniform transcendentality 4. Multiple zeta values are defined as in \cite{Lee:2011jt} and appear in the expansion, due to the last two master integrals.

At three loops the expression in terms of master integrals reads
\begin{align}
&\left\langle O_{0}(p) O_{0}(-p) \right\rangle^{(3)}=
\tfrac{8 (2 \epsilon -1)^3}{\epsilon ^3} \, \raisebox{-0.8mm}{\includegraphics[scale=0.1]{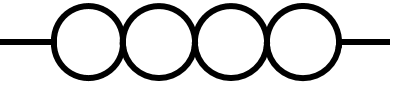}}
-\tfrac{4 (3 \epsilon -1) \left(408 \epsilon ^2+187 \epsilon +21\right) (2 \epsilon -1)^2}{\epsilon ^3 (4 \epsilon +1) (5 \epsilon +1)} \, \raisebox{-0.8mm}{\includegraphics[scale=0.1]{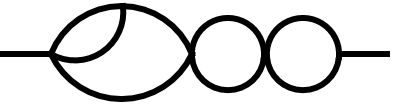}}
\nonumber\\&
+\tfrac{16 (4 \epsilon -1) \left(324 \epsilon ^2+157 \epsilon +18\right)(2 \epsilon -1)^2}{3 \epsilon ^3 (4 \epsilon +1) (5 \epsilon +1)}  \, \raisebox{-1.6mm}{\includegraphics[scale=0.1]{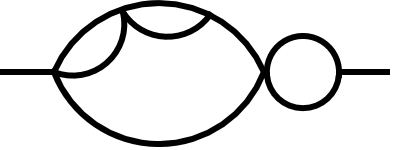}} 
-\tfrac{20 (5 \epsilon -1) (2 \epsilon -1)^2}{\epsilon ^3}  \, \raisebox{-2.5mm}{\includegraphics[scale=0.1]{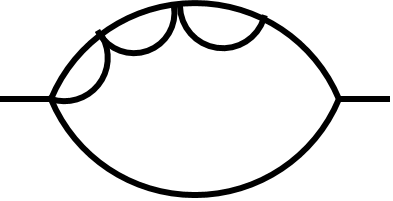}} 
\nonumber\\&
-\tfrac{4(3 \epsilon -1) (4 \epsilon -1) (5 \epsilon -1) \left(14484 \epsilon ^3+10379 \epsilon ^2+2433 \epsilon +190\right)}{9 \epsilon ^3 (3 \epsilon +1) (4 \epsilon +1) (5 \epsilon +1)}  \, \raisebox{-2.5mm}{\includegraphics[scale=0.1]{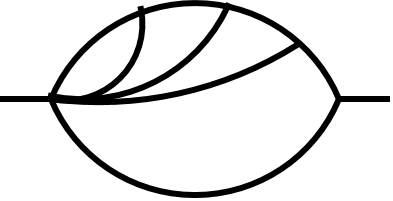}}   
-\tfrac{48 (4 \epsilon +1) (2 \epsilon -1)}{\epsilon  (5 \epsilon +1)}  \, \raisebox{-2.5mm}{\includegraphics[scale=0.1]{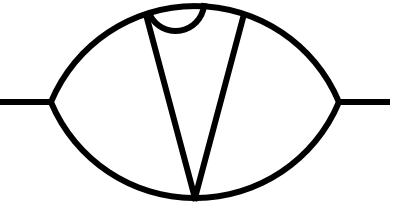}}
\nonumber\\&
+\tfrac{16 (3 \epsilon -1) (4 \epsilon -1) \left(1257 \epsilon ^3+833 \epsilon ^2+165 \epsilon +10\right) (2 \epsilon -1)}{9 \epsilon ^3 (3 \epsilon +1) (4 \epsilon +1) (5 \epsilon +1)}  \, \raisebox{-1.6mm}{\includegraphics[scale=0.1]{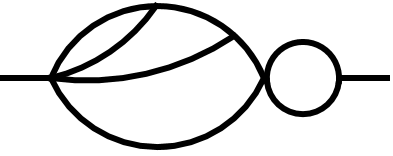}}
-\tfrac{48 (2 \epsilon -1)^2}{\epsilon ^2}  \, \raisebox{-1.6mm}{\includegraphics[scale=0.1]{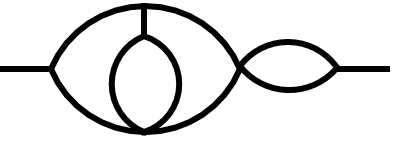}} 
\nonumber\\&
-\tfrac{4(3 \epsilon -1) (5 \epsilon -1) \left(3288 \epsilon ^3+2629 \epsilon ^2+579 \epsilon +38\right) (2 \epsilon -1)}{9 \epsilon ^3 (3 \epsilon +1) (4 \epsilon +1) (5 \epsilon +1)}  \, \raisebox{-2.5mm}{\includegraphics[scale=0.1]{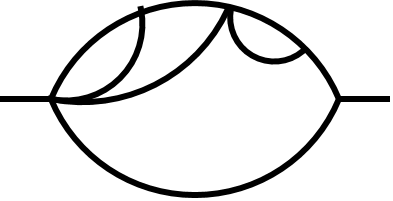}}
+\tfrac{60 (5 \epsilon -1) (2 \epsilon -1)}{\epsilon  (3 \epsilon +1)} \, \raisebox{-2.5mm}{\includegraphics[scale=0.1]{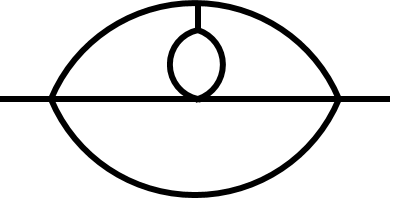}}
\nonumber\\&
+\tfrac{4(4 \epsilon -1) (5 \epsilon -1) \left(3592 \epsilon ^3+2938 \epsilon ^2+761 \epsilon +63\right) (2 \epsilon -1)}{3 \epsilon ^3 (3 \epsilon +1) (4 \epsilon +1) (5 \epsilon +1)}  \, \raisebox{-2.5mm}{\includegraphics[scale=0.1]{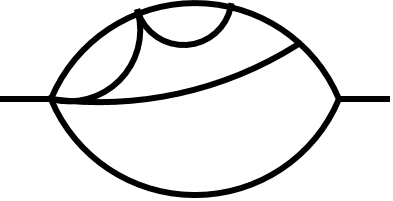}}
+\tfrac{16 (2 \epsilon -1)^2}{(3 \epsilon +1) (5 \epsilon +1)} \, \raisebox{-2.5mm}{\includegraphics[scale=0.1]{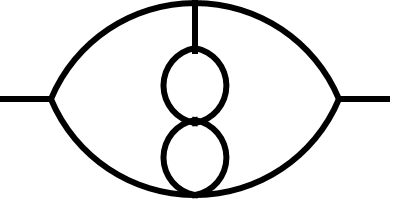}}
\nonumber\\&
+\tfrac{8 (3 \epsilon -1)^2 \left(329 \epsilon ^3+223 \epsilon ^2+49 \epsilon +4\right) (2 \epsilon -1)}{3 \epsilon ^3 (3 \epsilon +1) (4 \epsilon +1) (5 \epsilon +1)}  \, \raisebox{-0.8mm}{\includegraphics[scale=0.1]{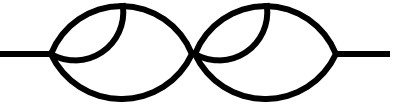}}
-\tfrac{12 (3 \epsilon -1) \left(128 \epsilon ^2+65 \epsilon +8\right) (2 \epsilon -1)}{\epsilon ^2 (4 \epsilon +1) (5 \epsilon +1)} \, \raisebox{-2.5mm}{\includegraphics[scale=0.1]{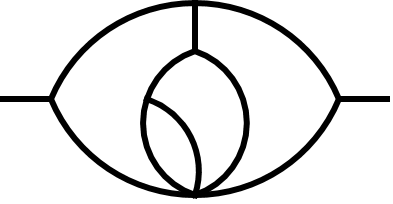}}
\nonumber\\&    
+\tfrac{32 \left(363 \epsilon ^2+205 \epsilon +26\right) (2 \epsilon -1)^2}{9 \epsilon ^2 (3 \epsilon +1) (5 \epsilon +1)} \, \raisebox{-2.5mm}{\includegraphics[scale=0.1]{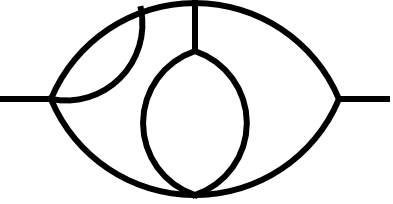}} 
-\tfrac{8 \left(300 \epsilon ^2+142 \epsilon +15\right) (2 \epsilon -1)^2}{3 \epsilon ^2 (4 \epsilon +1) (5 \epsilon +1)} \, \raisebox{-2.5mm}{\includegraphics[scale=0.1]{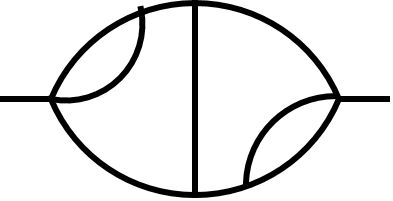}} 
+\tfrac{4 (2 \epsilon -1)}{\epsilon } \, \raisebox{-1.6mm}{\includegraphics[scale=0.1]{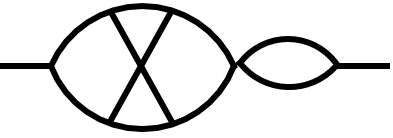}} 
\nonumber\\&  
+\tfrac{4 (2 \epsilon -1)}{\epsilon } \, \raisebox{-2.5mm}{\includegraphics[scale=0.1]{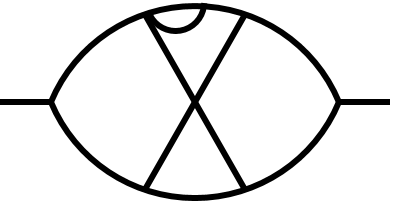}} 
-\tfrac{4 (6 \epsilon +1) (2 \epsilon -1)}{\epsilon  (4 \epsilon +1)} \, \raisebox{-2.5mm}{\includegraphics[scale=0.1]{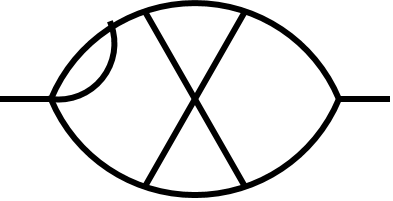}} 
-\tfrac{16 (4 \epsilon -1) (5 \epsilon -1) \left(104 \epsilon ^2+69 \epsilon +10\right) }{9 \epsilon ^2 (4 \epsilon +1) (5 \epsilon +1)} \, \raisebox{-2.5mm}{\includegraphics[scale=0.1]{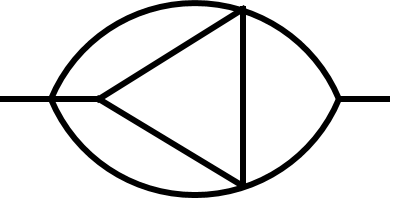}} 
\nonumber\\&
+\tfrac{20 (5 \epsilon -1) \left(19 \epsilon ^2+13 \epsilon +2\right)}{3 \epsilon  (3 \epsilon +1) (5 \epsilon +1)} \, \raisebox{-2.5mm}{\includegraphics[scale=0.1]{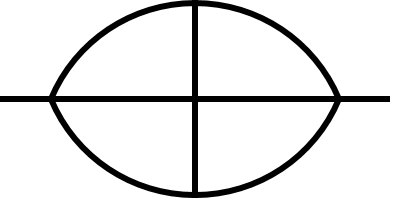}} 
-\tfrac{8 \epsilon}{(4 \epsilon +1) (5 \epsilon +1)} \, \raisebox{-2.5mm}{\includegraphics[scale=0.1]{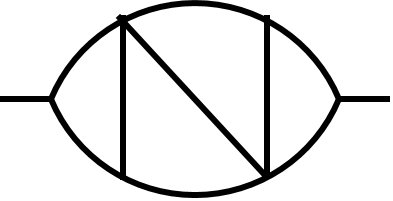}} 
-\tfrac{16 (6 \epsilon -1) (6 \epsilon +1)}{(4 \epsilon +1) (5 \epsilon +1)} \, \raisebox{-2.5mm}{\includegraphics[scale=0.1]{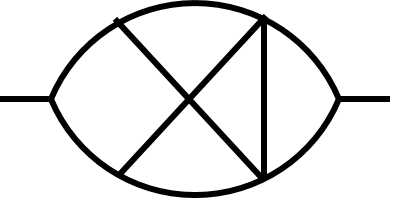}} 
\nonumber\\&
-\tfrac{2 (3 \epsilon +1)}{5 \epsilon +1} \, \raisebox{-2.5mm}{\includegraphics[scale=0.1]{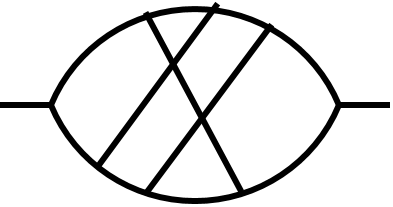}} 
-\tfrac{2 (3 \epsilon +1) (4 \epsilon +1)}{\epsilon  (5 \epsilon +1)} \, \raisebox{-2.5mm}{\includegraphics[scale=0.1]{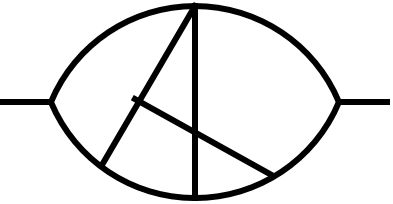}} 
+\tfrac{(3 \epsilon +1)^2 }{\epsilon  (5 \epsilon +1)} \, \raisebox{-2.5mm}{\includegraphics[scale=0.1]{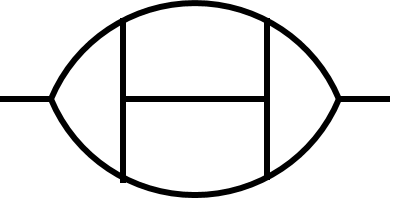}} 
\end{align}
Replacing the expansions of the integrals given in \cite{Lee:2011jt}, or letting Forcer do the job, yields the result
\begin{align}\label{eq:3L}
    &n^{(3)}=-980 \zeta _7 \epsilon+\left(-5560 \zeta _3 \zeta _5-3430 \zeta _8\right) \epsilon ^2    
    +\left(-\tfrac{17560}{3} \zeta _3^3-13900 \zeta _6 \zeta _3-8340 \zeta _4 \zeta _5
\right.\nonumber\\&\left.    
    +1470 \zeta _2 \zeta _7-\tfrac{292220}{9} \zeta _9\right) \epsilon ^3
    +\left(\tfrac{18330}{7} \zeta_{7,3}-26340 \zeta _4 \zeta _3^2+8340 \zeta _2 \zeta _5 \zeta _3-76170 \zeta _7 \zeta _3-\tfrac{391050}{7} \zeta _5^2
\right.\nonumber\\&\left.    
    -\tfrac{4537861}{28} \zeta _{10}\right)\epsilon ^4 
    +\left(64368 \zeta _3 \zeta_{5,3}-77152 \zeta_{5,3,3}
    +8780 \zeta _2 \zeta _3^3+52040 \zeta _5 \zeta _3^2-384846 \zeta _8 \zeta _3
\right.\nonumber\\&\left.    
    +\tfrac{40041999}{8} \zeta _{11}-\tfrac{179855}{2} \zeta _5 \zeta _6-\tfrac{10269410 }{3}\zeta _2 \zeta _9-\tfrac{1616769}{4} \zeta _4 \zeta _7\right)\epsilon ^5 
    +\left(70184 \zeta _4 \zeta_{5,3}+\tfrac{1223545}{7} \zeta _2 \zeta_{7,3}
\right.\nonumber\\&\left.    
    -\tfrac{178720}{3} \zeta_{6,4,1,1}-\tfrac{1569410}{9} \zeta_{9,3}+\tfrac{1620880}{9} \zeta _3^4+\tfrac{218205}{2} \zeta _6 \zeta _3^2-\tfrac{2257405}{6} \zeta _4 \zeta _5 \zeta _3
    +\tfrac{2129965}{3} \zeta _2 \zeta _7 \zeta _3
\right.\nonumber\\&\left.    
    -\tfrac{39914030}{27} \zeta _9 \zeta _3+\tfrac{6138365}{21} \zeta _2 \zeta _5^2-\tfrac{1883141}{3} \zeta _5 \zeta _7-\tfrac{400117736081}{66336} \zeta _{12}\right)\epsilon ^6 + {\cal O}(\epsilon^7)
\end{align}
which again obeys uniform transcendentality 6 up to the order to which the expansion was performed.
These findings motivate the following conjecture:
\begin{conjecture}
The quantum corrections to the two-point function of protected bi-scalar operators in ${\cal N}=4$ SYM have uniform degree of transcendentality $2l$ at loop $l$.
In particular, the leading order $\epsilon$ term at loop $l$ displays transcendentality degree $(2l+1)$.
\end{conjecture}
One might object that such a statement is ill-defined or trivial. One can always modify the normalization of an operator and, for instance, cook up a factor cancelling out lower transcendental contributions perturbatively in $\epsilon$ in the two-point function.
Our point is that, remarkably, for these operators such a redefinition is unnecessary and the two-point function of the bare $O_0$ operators naturally produces, with the adopted regularization scheme, a result displaying uniform transcendentality on the spot.

Requiring that the final result is subleading in $\epsilon$ and uniformly transcendental implies certain relations between the coefficients of the master integrals in terms of which it can be expressed. However, such constraints are generically not sufficient for fixing them completely, unless some additional assumptions are made on the form of the rational functions multiplying master integrals. Moreover, even the most efficient ansatz minimizing the number of undetermined coefficients would at best determine all up to an overall factor, which ought to be fixed independently of uniform transcendentality. In conclusion, we do not believe this conjecture could be leveraged to bootstrap the two-point functions at higher loops, unless complemented with other techniques.

\section{Higher perturbative orders}

Assuming uniform transcendentality holds to all loops, we could wonder if some extrapolation to higher orders is possible.
Starting from four-loops, spinning twist-two operators develop non-planar corrections to their anomalous dimensions. It is reasonable to expect that the $\epsilon$ sub-leading two-point function of protected operators also receives non-planar corrections. This would allow for more coefficients in the expansion \eqref{eq:perturbative}.
Ignoring this possible issue or focussing on planar corrections, we could speculate whether the leading contributions at order $\epsilon$, which look particularly simple in \eqref{eq:1L}, \eqref{eq:2L} and \eqref{eq:3L}, could be extrapolated from the very few coefficients at our disposal: $(-12\zeta_3,100\zeta_5, -980\zeta_7,\dots)$. 
According to the OEIS \cite{OEIS}, the coefficients of the zeta values could be the first few terms of the sequence \href{https://oeis.org/A000888}{A000888}. Accordingly, we put forward the following conjecture
\begin{conjecture}
The leading order $\epsilon$ quantum correction to two-point functions of protected scalar bi-linear operators in ${\cal N}=4$ SYM at loop order $l$ is
\begin{equation}
\frac{\left\langle O_{0}(p) O_{0}(-p) \right\rangle^{(l)}}{\left\langle O_{0}(p) O_{0}(-p) \right\rangle^{(0)}} = (-1)^l\,\frac{(2l+2)!^2}{(l+2)(l+1)!^4}\, \zeta_{2 l+1} \, \epsilon + {\cal O}\left(\epsilon^2\right)
\end{equation}
\end{conjecture}
This is a bold extrapolation only based on three non-trivial data points. Still, such sort of expression looks like a reasonable combinatorial factor that can emerge in Feynman integrals numerology.
Taking the sequence seriously, it would predict the four- and five-loop corrections at order $\epsilon$ to read
\begin{align}
\frac{\left\langle O_{0}(p) O_{0}(-p) \right\rangle}{\left\langle O_{0}(p) O_{0}(-p) \right\rangle^{(0)}} &\overset{?}{=} 1+\left(-12 \zeta_3\lambda + 100 \zeta_5\lambda^2 - 980 \zeta_7\lambda^3 + 10584 \zeta_9\lambda^4\right.\nonumber\\&\left. - 121968 \zeta_{11}\lambda^5 
+ {\cal O}\left(\lambda^6\right)\right)\epsilon + {\cal O}\left(\epsilon^2\right) \overset{?}{+} {\cal O}\left(N^{-1}\right)
\end{align}
though it is not clear to us what the role of possible non-planar contributions could be.
For this reason, and due to the feeble support we have gathered in favor of such a conjecture, it would be interesting to verify this prediction via a direct computation at the next loop order, four. The relevant five-loop master integrals are in principle available \cite{Georgoudis:2018olj,Georgoudis:2021onj} up to transcendental weight 9.

\section{Integrated three-point functions}

The discovery of the uniform transcendentality property of two-point functions of twist-two scalars stemmed from a three-point function calculation.
The latter was performed integrating three-point functions over the position of one of the operators, in order to extract the structure constant from a two-point problem \cite{Plefka:2012rd}. In momentum space, that translates into a soft limit for one of the operators' momenta, effectively reducing the calculation from a three-point function to a two-point one.
Again, quantum corrections to three-point functions involving three protected twist-two scalars are expected to vanish in four dimensions, implying that the integrated three-point function at loop level is subleading in $\epsilon$.
This is indeed what happens and it was used in \cite{Bianchi:2023cbc} as a consistency check.

Inspecting the actual results for the three-point functions with a soft limit, reveals again rather simple transcendental values at low orders in $\epsilon$.
The result breaks uniform transcendentality, however the source of non-uniformity is a simple loop dependent overall factor.
This motivates exploring the possibility that some pattern might appear in this computation as well.
Up to three loops, the result reveals an interesting structure
\begin{align}\label{eq:3ptsoft}
\left\langle O_{0}(p) O_{0}(-p) O_{0}(0) \right\rangle &= (-1+2\epsilon)\left\langle O_{0}(p) O_{0}(-p) \right\rangle^{(0)}
+(-1+4\epsilon)\left\langle O_{0}(p) O_{0}(-p) \right\rangle^{(1)}\lambda \nonumber\\&
+(-1+6\epsilon)\left\langle O_{0}(p) O_{0}(-p) \right\rangle^{(2)}\lambda^2
+(-1+8\epsilon)\left\langle O_{0}(p) O_{0}(-p) \right\rangle^{(3)}\lambda^3\nonumber\\&
+{\cal O}\left(\lambda^4\right)
\end{align}
This can be proven exactly at the level of the master integrals above and is a non-trivial result which only emerges summing over all Feynman diagrams. In fact, if perhaps the uniform transcendentality property of two-point functions might not have been completely surprising, this finding looks interestingly bizarre.
Accordingly, we formulate a third conjecture:
\begin{conjecture}
The soft limit of a three point-function of protected twist-two operators in ${\cal N}=4$ SYM at loop order $l$ is equivalent to the corresponding two-point function multiplied by $\left(-1+(2l+2)\epsilon\right)$
\begin{equation}\label{eq:3ptsoftconj}
\left\langle O_{0}(p) O_{0}(-p) O_{0}(0) \right\rangle^{(l)} = \left(-1+(2l+2)\epsilon\right)\left\langle O_{0}(p) O_{0}(-p) \right\rangle^{(l)}
\end{equation}
\end{conjecture}
The objects we are computing do not correspond to any physical observable of interest, though. In particular they are not related to $\epsilon$ corrections to structure constants, which indeed sounds like a meaningless concept. They are just soft limits of a three-point calculation in momentum space.

A natural question arises whether there is a way of connecting the soft limit of three-point functions to two-point functions in such a fashion to expose and prove the property highlighted above.
In practice, the presence of a third operator amounts in this specific calculation to dressing the scalar propagators of the two-point function diagrams with an extra power, in all possible ways and taking into account symmetries. This operation can be thought of as the action of some differential or index raising operator, typical of IBP identities, on certain propagators of the relevant Feynman integrals. In fact, in our calculation such extra powers are dealt with via such a technique in an automated fashion, which, albeit convenient, obscures the details of the process.
The way this index raising operation acts on each diagram and combines across different graphs to produce \eqref{eq:3ptsoft} remains rather mysterious to us.
In fact, such a property implies certain constraints between the coefficients of the various diagrams entering the two-point function calculation, which, however, are not sufficient nor convenient for attempting some kind of bootstrap.

\section{Conclusions}

In this note we pointed out that two-point functions of protected bi-linear scalar operators in ${\cal N}=4$ SYM at loop order $l$ exhibit uniform degree of transcendentality $2l$. The coefficient of the leading term was conjectured to all orders and a surprising perturbative relation was uncovered with three-point functions where one of the momenta vanishes.

What is the origin of the uniform transcendentality property uncovered in this note? Two- and higher-point correlation functions can be constructed via unitarity from form factors \cite{Engelund:2012re}. Form factors of protected operators in ${\cal N}=4$ SYM in turn also display uniform transcendentality of their numeric values \cite{Gehrmann:2011xn,Huber:2019fxe,Agarwal:2021zft,Lee:2021lkc} or the functions of the momenta invariants \cite{Bork:2010wf,Brandhuber:2012vm,Brandhuber:2014ica,Banerjee:2016kri,Lin:2020dyj,Dixon:2020bbt,Dixon:2022rse,Dixon:2022xqh}.
This suggests that the transcendentality property presented in this note might be inherited from that. However, the precise way through which uniform transcendentality could be transferred when gluing form factors into correlators remains unclear to us. The uniform transcendentality of form factors, or of scattering amplitudes, is also more a heuristic finding rather than a proven fact. Hence, even if a precise connection with form factors could be established, it would only partially explain why these two-point functions exhibit uniform transcendentality.

Finally, we comment on possible developments.
Higher-point correlators: if we integrate over the position of $(n-2)$ bi-scalar operators of an $n$-point correlation function would some relation akin to \eqref{eq:3ptsoft} emerge? It should be possible to check this again up to three-loop order with the same techniques implemented in \cite{Bianchi:2023cbc}.

Higher dimension BPS operators: do their two-point functions in dimensional regularization exhibit uniform transcendentality too? It is worth inspecting. Up to two loops it is not an intricate calculation, since only momentum integrals with up to four loops appear for which the technology already used in \cite{Bianchi:2023cbc} is available.

ABJM: could two-point functions of protected twist-one operators in ABJM display uniform transcendentality? To finite order in $\epsilon$ they do at leading order in the coupling \cite{Minahan:2009wg,Young:2014lka,Bianchi:2020cfn} and to all orders in a certain color limit \cite{Bianchi:2016rub}. Extending the two-loop result to higher powers in the $\epsilon$ expansion should be possible, the only obstacle being expanding the non-planar master integral to higher orders in $\epsilon$, which should be doable. It has probably already been done, at least numerically, though we could not find it published. Pushing the analysis to the next non-trivial perturbative order would require a four-loop computation, though. No results for the corresponding five-loop propagator integrals in three dimensions is available.

\acknowledgments

This work was supported by Fondo Nacional de Desarrollo Cient\'ifico y Tecnol\'ogico, FONDECYT 1220240. We also thank Beca Santander de Movilidad Internacional, for funding a research visit at Niels Bohr Institute and Icelandic University, where most of this work was performed. We thank these institutions for hospitality and especially our hosts Charlotte Kristjansen and Valentina Giangreco Marotta Puletti.

\vfill
\newpage

\bibliographystyle{JHEP}
\bibliography{biblio3}

\end{document}